\documentclass[preprintnumbers]{revtex4-2}

\usepackage{graphicx}
\usepackage{dcolumn}
\usepackage{bm}
\usepackage[utf8]{inputenc}
\usepackage{tabularx}
\usepackage{array}
\usepackage{graphics}
\graphicspath{{figures/}}
\usepackage{graphicx}
\usepackage{psfrag}
\usepackage{epsfig}
\usepackage{amsmath}
\usepackage{amssymb}
\usepackage[normalem]{ulem}
\usepackage{setspace}
\usepackage{rotating}
\usepackage{colortbl}
\usepackage{tabularx}
\usepackage{tikz-feynman}
\usepackage{slashed}
\usepackage{enumitem}
\makeatletter
\usepackage{textcomp}
\usepackage{caption}
\usepackage{subcaption}
\usepackage[hidelinks]{hyperref}
\hypersetup{%
  colorlinks=true,
  allcolors=blue
}
\usepackage{cleveref}

\newcommand{\be}{\begin{equation}}
\newcommand{\ee}{\end{equation}}
\newcommand{\bea}{\begin{equation} \begin{aligned}}
\newcommand{\eea}{\end{aligned} \end{equation}}	
	
\definecolor{niceblue}{rgb}{0.388235, 0.627451, 0.847059}

\begin{document}

\preprint{FERMILAB-PUB-24-0119-T}

\title{Could SBND-PRISM probe Lepton Flavor Violation?}

\author{Gustavo F. S. Alves}
 \email{gustavo.figueiredo.alves@usp.br}
 \author{Renata Zukanovich Funchal}
 \email{zukanov@if.usp.br}
\affiliation{%
Departamento de F\'{\i}sica Matem\'atica, Instituto de F\'{\i}sica\\
Universidade de S\~ao Paulo, 05315-970 S\~ao Paulo, Brasil
}%

\author{Pedro A. N. Machado}
\email{pmachado@fnal.gov}
\affiliation{Particle Theory Department, Fermi National Accelerator Laboratory, Batavia, IL 60510, USA}

\date{\today}

\begin{abstract}

We investigate the possibility of using the Short-Baseline Near Detector (SBND) at Fermilab to constrain lepton flavor violating decays of pions and kaons. 
We study how to leverage SBND-PRISM, the use of the neutrino beam angular spread to mitigate systematic uncertainties, to enhance this analysis.
We show that SBND-PRISM can put stringent limits on the flavor violating branching ratios $\rm{BR}(\pi^+ \to \mu^+ \nu_e) = 8.9 \times 10^{-4}$, $\rm{BR}(K^+ \to \mu^+ \nu_e) = 3.2 \times 10^{-3}$, improving previous constraints by factors 9 and 1.25, respectively. 
We also estimate the SBND-PRISM sensitivity to lepton number violating decays, $\rm{BR}(\pi^+ \to \mu^+ \overline{\nu}_e)= 2.1 \times 10^{-3}$ and  $\rm{BR}(K^+ \to \mu^+ \overline{\nu}_e) = 7.4 \times 10^{-3}$, though not reaching previous BEBC limits.
Last, we identify several ways how the SBND collaboration could  improve this analysis.\footnote{The work and conclusions presented in this publication are not to be considered as results from the SBND collaboration.}

\end{abstract}

\maketitle

\section{Introduction}
\label{sec:intro}

Neutrino flavor oscillations represent to this day the sole 
evidence of lepton flavor violation (LFV). They also 
may be connected, if neutrinos 
are massive Majorana fermions, to lepton number violation (LNV). 
The lack of any conclusive 
signals from the Large Hadron Collider experiments  triggers 
the interest on low energy precision measurements.
Besides 
the classic searches
for LFV processes involving charged leptons such as  
$\mu \to e \gamma$, 
$\mu \to 3 e$ or $\mu\to e$ -- conversion in nuclei, and   for  
the quintessential LNV 
neutrinoless double-$\beta$ decay, meson 
decays  can also play a role in scrutinizing these symmetries.

In fact, light pseudoscalar  meson  leptonic decays,   the main sources of the conventional neutrino fluxes in accelerator neutrino oscillation experiments, have already been used to this purpose in the past.
The best bounds  come from the Big European Bubble Chamber (BEBC)~\cite{Cooper-Sarkar:1981bam, Lyons:1981xs, Workman:2022ynf}, and are, correspondingly, 
$\rm{BR}(\pi^+ \to \mu^+ \nu_e) < 8.0 \times 10^{-3}$, $\rm{BR}(K^+ \to \mu^+ \nu_e) < 4 \times 10^{-3}$ for the lepton number conserving (LNC) and $\rm{BR}(\pi^+ \to \mu^+ \overline{\nu}_e) < 1.5 \times 10^{-3}$, $\rm{BR}(K^+ \to \mu^+ \overline{\nu}_e) < 3.3 \times 10^{-3}$ for the LNV modes. 
Moreover, the combination of very intense beams and highly sensitive detectors, 
now available at  some   
particle accelerator facilities,
allows to use them once 
again as important probes of 
these Beyond the Standard Model (BSM)
manifestations of new physics. 

From the theoretical point of view LNV and LFV contributions 
to rare meson decays are expected to  be connected to UV complete models with new degrees of freedom appearing above the electroweak scale. 
A useful way to tackle the problem is by means of the effective field theory approach. One constructs an  effective Lagrangian consisting of the 
SM part and a series of all possible non-renormalizable  operators 
of dimension $d \geq 5$ invariant under $SU(3)_c\times SU(2)_L \times U(1)_Y$ involving the 
SM fields 
\begin{equation}
{\cal L}_{\rm eff} = {\cal L}_{\rm SM} + \sum_i \sum_{d \geq 5} \frac{c_i^{(d)}}{(\Lambda^{(d)})^{d-4}} {\cal O}_i^{d}\, ,
\end{equation}
where $c_i^{(d)}$ is the Wilson coefficient of operator $\mathcal{O}_i^d$, which will take it to be 1 for simplicity, and $\Lambda^{(d)}$ is the BSM scale.
The lowest dimension of these operators ($d=5$) is the well-known LNV Weinberg operator~\cite{Weinberg:1979sa} that lead to Majorana neutrino masses. 
There are four (two vector and two scalar) $d=6$ quark-lepton charged current operators that may induce LFV, 
either conserving or violating lepton number, in
light meson leptonic decays~\cite{Li2020EFTforLNVandLFV}. 
Allowing for a single of these operators to saturate the experimental bound one  can probe a BSM  
scale $\Lambda^{(6)}$ up to $(0.6 - 4.7)$ TeV using 
$\pi^+ \to \mu^+ \nu_e (\bar \nu_e)$  or up to $(1.3 - 7.7)$
TeV using $K^+ \to \mu^+ \nu_e (\bar \nu_e)$. 
In Ref.~\cite{Babu:2001ex} 
a classification of $\Delta L=2$  operators with $d<12$ have been provided. 
Using this classification 
the authors of  
Ref.~\cite{Deppisch2020EFTforLNVandLFV} have investigated 
constraints on  LNV operators using rare kaon decays. In particular,  
they have shown that the current experimental limits on $\pi^+ (K^+) \to \mu^+ \bar \nu_e$ can discard 
$\Lambda^{(7)} \lesssim 1.9 \, (2.4)$
TeV. 

There have been several analyses on LFV and LNV connected to hadronic decays using the EFT formalism; for example Refs.~\cite{Bhattacharya:2015vja, Bischer:2019ttk} studied the impact on LFV of adding right-handed neutrinos to the standard model EFT, and Refs.~\cite{Bhattacharya:2011qm, Cirigliano:2013xha} estimated the  beta and pion decays.
Some of the operators discussed above can also
be realized in several 
UV complete models, such as 
supersymmetric 
R-parity violating models~\cite{Hall:1983id} or  leptoquark models~\cite{Davidson1994UVcompleteModelLNVandLFV,Babu:1994kb,Babu:1995uu,
Deppisch2020EFTforLNVandLFV}.
Additionally, Refs.~\cite{Babu:2002ica, Babu:2016fdt} have discussed how UV complete scenarios of LFV can affect muon decays and alleviate the LSND anomaly.
For a comprehensive study on how to relate neutrino mass mechanisms and LFV interactions, see Ref.~\cite{Jana:2022aut}.
Note that the Lepton Flavor Universality observables defined for  pseudoscalar mesons decays as 
\begin{equation}
R^{M}_{e/\mu} \equiv \frac{\Gamma(M^+ \to e^+ + \rm invisible)}{\Gamma(M^+ \to \mu^+ + \rm invisible)}\, , \quad \quad  M= \pi,K \; ,
    \label{eq:ratios}
\end{equation}
allow to obtain from data and SM calculations the double ratios $R^{\pi}_{e/\mu}[\rm exp]/ R^{\pi}_{e/\mu}[\rm SM] = 0.9994\pm 0.0024$ and
$R^{K}_{e/\mu}[\rm exp]/ R^{K}_{e/\mu}[\rm SM] = 1.0044\pm 0.0041$~\cite{Cirigliano:2007xi}. The experimental ratios are, however, not very sensitive to the type of LNV or LFV  processes we have discussed above as the final state neutral particle/antiparticle  nature and flavor are not determined.

The Short Baseline Near Detector (SBND) is an ideal candidate to search for 
these type of processes. It will collect unprecedented statistics of neutrino events and set the beam content by probing the unoscillated neutrino flux. This suppresses natural contamination of $\nu_e$ from neutrino oscillations, mitigating background for the flavor violating meson decays searches~\footnote{Note that the baseline of SBND (110 m) is much shorter than the shortest neutrino oscillation length 
$L_{\rm osc}^{\rm atm}=(4 \pi E/\Delta m^2_{31}) \sim 2 \times 10^3 \; \rm km $, so there is no contribution from neutrino oscillation to the flux.
}. 
Before proceed with more details, we can make a simple estimate of the sensitivity of SBND in constraining light meson decays to the wrong neutrino. 
If we consider $M\to\mu\nu_e$, we can compare the BSM $\nu_e$ flux with the uncertainty in the intrinsic $\nu_e$ contamination, namely,
\begin{equation}
    \text{BR} (M^+ \to \mu^+ \nu_e) \phi_{\nu_\mu}^M = \sigma_{e}\phi_{\nu_e},
    \label{eq:toy_pion_flav_viol}
\end{equation}
where $\phi_{\nu_\alpha}^M$ is the beam flux of $\nu_\alpha$ coming from the decays of meons $M$ and $\sigma_e$ is the uncertainty in the  electron neutrino flux for SBND. 
For a significant part of the neutrino spectrum at SBND $\phi_{\nu_\mu} \sim \mathcal{O}(100)\ \phi_{\nu_e}$.
We take a $\sigma_\phi \sim 10\%$  uncertainty on the $\nu_e$ contamination~\cite{MicroBooNE:2018efi}.
This means that a  $\text{BR}(M \to \mu \nu_e) \sim \mathcal{O}(10^{-3})$ could be achieved, demonstrating that SBND can potentially provide competitive constraints on LFV pion decays.

Building from this simple argument, we also note that the SBND collaboration has identified a potential use angular spread of the Booster Neutrino Beam --the SBND-PRISM technique~\cite{delTutto2021SBND}, similar to previous studies~\cite{nuprism, dunetdr}.
SBND can measure neutrino events in off-axis positions by identifying the vertex of the event.
Since the distance to the target is about 110 meters, compared to SBND largest dimension of 5 meters, the nontrivial angular spread translates into specific $\nu_\mu$ and $\nu_e$ energy spectra that are off-axis angle dependent.
Therefore, as the ratio of $\nu_\mu$ to $\nu_e$ events change with the off-axis angle, we can leverage that as an additional handle to control background uncertainties. 
In this work we aim to show how SBND can improve the searches for these exotic $\pi$ and $K$ decays. 

In  \cref{sec:sbnp} we discuss the details of the SBND experiment and the PRISM technique that are pertinent for our study. 
We discuss how we simulate the signal and our statistical procedure in \cref{sec:analysis}, followed by our results in  \cref{sec:results}. 
To conclude,  \cref{sec:conclusions} summarizes our key findings. We also discuss potential ways to improve the experimental sensitivity beyond what we have considered.

\section{The Short Baseline Neutrino Program}
\label{sec:sbnp}
The Booster Neutrino Beam (BNB) line at Fermilab provides the neutrino flux for the Short Baseline Neutrino (SBN) program~\cite{Machado:2019oxb}. 
This is achieved by colliding a beam of 8 GeV protons with a beryllium target, producing a secondary beam of hadrons. 
In sequence, magnetic horns select the hadron charge allowing it to function in two different modes: the neutrino mode, where the positively charged hadrons are focused and the antineutrino mode where negatively charged hadrons are focused instead. 
The selected flux of particles propagate in the 50~m long decay pipe giving rise to the neutrino beam through their subsequent decays. 
The SBN program is expected to run mostly in the neutrino mode where the flux is mainly comprised of $\nu_\mu$ (96\%) originated predominantly from $\pi^+ \to \mu^+ \nu_\mu$ for energies up to approximately $2\ \rm GeV$, beyond which $K^+ \to \mu^+ \nu_\mu$  decays become the dominant source of $\nu_\mu$ production.
The BNB flux contains  a small fraction of $\overline{\nu}_\mu$ ($5\%$), stemming from the contamination of wrong-sign hadrons and muon decays. 
Last, the $\nu_e$ and $\overline{\nu}_e$ beam contamination of about 0.5\% originates from both kaon and muon decays.

The SBND detector is the first experiment in the BNB line, located 110 m away from the BNB target. 
It is a 112 ton liquid argon time projection chamber (LArTPC) with an active volume of 5.0 m (L) $
\times$ 4.0 m (W) $\times$ 4.0 m (H). With an initially planned exposure of $6.6 \times 10^{20}$ protons on target (POT), SBND will collect about 7 million $\nu_\mu$ charged current (CC) interactions and 50 thousand $\nu_e$ CC interactions, in a three years period.
The LArTPC technology can provide a 3D reconstruction of neutrino-argon interactions events by reconstructing the tracks of charged particles as they travel throught the detector ionizing the medium.

\begin{widetext}
        \begin{figure}[t]
            \includegraphics[width = 0.7\textwidth]{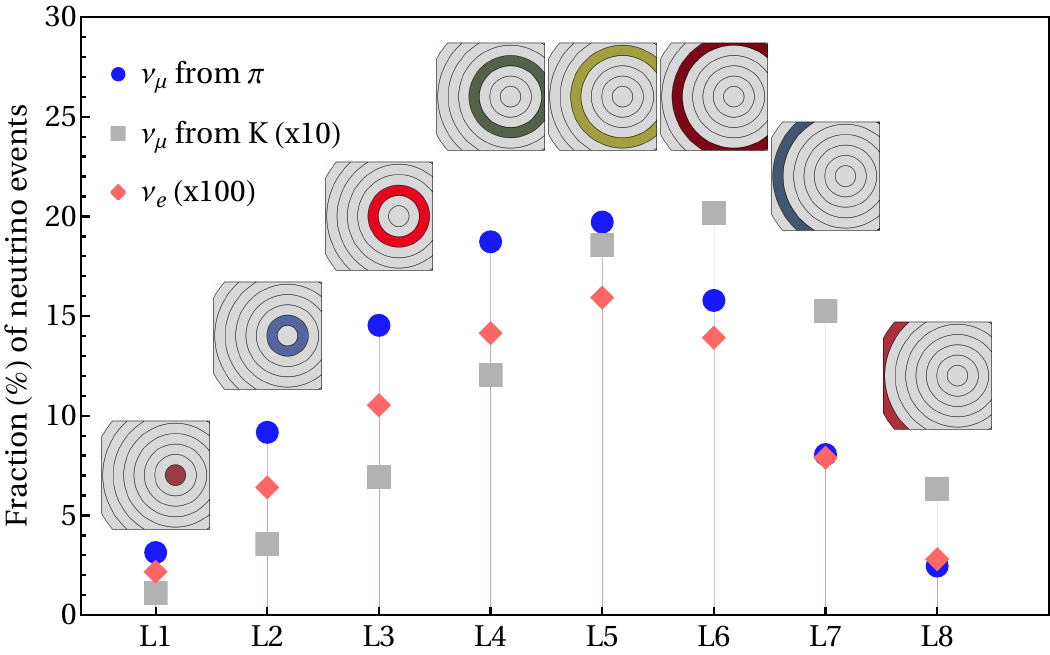}
            \caption{Fraction of CC events for $\nu_\mu$ from $\pi$ decay (blue), $\nu_\mu$ from $K$ decay (gray) and intrinsic $\nu_e$ (salmon) expected at  each layer labeled L1 to L8 of SBND-PRISM. The insets pictured above each fraction represent the front view of the SBND detector, segmented in eight layers, in which the corresponding layer is highlighted in color. 
            }
            \label{fig:evt_prof_SBND}
        \end{figure}
\end{widetext}

The term SBND-PRISM denotes the experiment's capability to conduct measurements at off-axis positions. 
The detector can be logically segmented into eight slices, each corresponding to an increasing off-axis angle within the range of $[0^\circ,1.6^\circ]$. 
The kinematics of two- and three-body meson decays dictate that the energy spectra of $\nu_\mu$ coming from $\pi\to\mu\nu$ tend to become narrower and shifts to lower energies as we move off-axis. 
In contrast, the $\nu_e$ from kaons and muons three-body decays are significantly less affected by the off-axis angle. 
This distinction implies a layer-dependent profile in the distribution of $\nu_\mu$ and $\nu_e$ events~\cite{delTutto2021SBND}, which can be leveraged to mitigate systematic uncertainties, in particular those related to neutrino-nucleus cross sections.
In Fig.~\ref{fig:evt_prof_SBND} we 
present the fraction of events for $\nu_\mu$ from $\pi$ decays, 
$\nu_\mu$ from $K$ decays, and intrinsic $\nu_e$ expected at each layer. 
In the next section we will describe in detail how we extracted these fractions.

\section{Analysis}
\label{sec:analysis}

In order to study the SBND capability to 
improve the constraints on LFV 
meson  decays, we rely on the distribution of 
simulated events and fluxes 
publicly provided by the SBND collaboration~\cite{delTutto2021SBND}, for $6.6\times 10^{20}$ POT. 
This  information allows us 
to know the total count of $\nu_\mu$ and $\nu_e$ CC events as well as their energy spectrum at SBND for each off-axis layer. 
However, it lacks the parent particle information, crucial for our purposes.
In the Booster Neutrino Beam, muon neutrinos below $E_{\rm flip}$ of a few GeV are dominantly produced by pion decays, while above $E_{\rm flip}$ neutrino production via kaon decay dominates~\cite{MiniBooNE:2008hfu}.
This flip between parent particle is fairly sharp.
Therefore, to estimate the corresponding  $E_{\rm flip}^i$ for each layer $i$, we do the following.

We simulate an 8 GeV proton beam with \texttt{PYTHIA}~\cite{bierlich2022}, generating the meson flux.
To mock the magnetic horn effect, we align all positive mesons in the beam direction, and decay them to neutrinos.
First, we select the neutrinos that cross the front face of SBND, and calculate at which energy the neutrino flux from pion decay cross the flux from kaon decays.
This yields $E_{\rm flip}=2.35$~GeV in comparison with the official BNB simulation $E_{\rm flip}^{\rm official}\simeq 2.3$~GeV~\cite{MiniBooNE:2008hfu}, which serves to validate this approach as a decent zero order approximation.
Then we repeat this procedure applying angular cuts to select the neutrinos that cross each layer $i$, obtaining $E_{\rm flip}^i$.
For simplicity, we use $E_{\rm flip}^i$ as a sharp cut to define the parent particle, that is, we approximate that all muon neutrinos below $E_{\rm flip}^i$ come from pion decays, while above that energy come from kaon decays.
Since we will not explicitly use spectral information, we do not expect the sharp cut to impact our analysis in any significant way.

\begin{figure}[t]
    \centering
    \includegraphics[width = \textwidth]{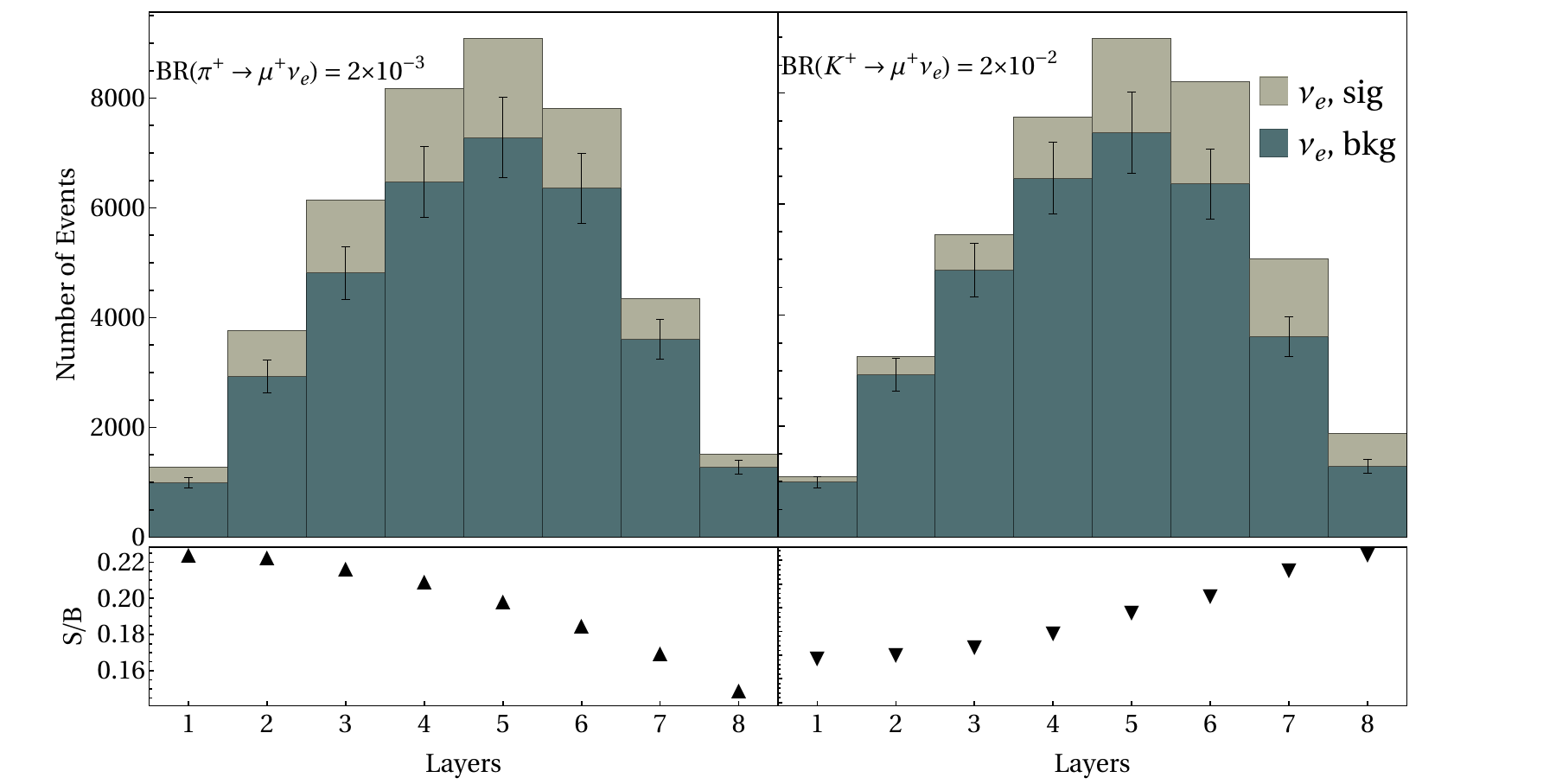}
    \caption{Number of signal (gray) and background(dark green) $\nu_e$ events, in each  SBND-PRISM layer for two representative branching ratios of LFV pion (left panel) and kaon (right panel) decays. For reference, we show a 10\% fully correlated uncertainty error bar.
    In the lower panels, we show the signal-to-background ratio $S/B$.
    }
    \label{fig:evts_vs_bkg}
\end{figure}

With that information at hand, we can estimate our signal coming from the flavor violating pions and kaon decays.
Since the two-body pion and kaon decay is dominated by the branching ratio into $\mu\nu_\mu$, we can obtain our signal flux $\phi_{\nu_e,{\rm sig}}^i$ by multiplying the flux of muon neutrinos coming from the decays of meson $M$ by the nonstandard branching, that is, 
\begin{equation}
    \phi_{\nu_e,{\rm sig}}^i = \phi_{\nu_\mu}^{M,i}\,{\rm BR}(M\to\mu\nu_e),
\end{equation}
where $i$ denotes the SBND-PRISM layer.
The key of the impact of SBND-PRISM lies in the different SBND-PRISM layer dependence of the signal and background fluxes, as the signal originates in two-body decays but the  intrinsic $\nu_e$ background comes from three-body decays.
To make this point more clear, we present in \cref{fig:evts_vs_bkg} the number of events for signal (gray) and background (dark green) in each layer for two representative values of the nonstandard LFV decays: ${\rm BR}(\pi^+\to\mu^+\nu_e)=2\times 10^{-3}$ (left panel) and ${\rm BR}(K^+\to\mu^+\nu_e)=2\times 10^{-2}$ (right panel).
For reference, we show a 10\% fully correlated systematic error in the prediction, as it will be discussed later.
We also display in the lower panels the signal-to-background ratio $S/B$ for these benchmarks.
We see clearly that while for nonstandard pion decays $S/B$ decreases 
with opening angle, while for kaon decays the opposite happens. 
By leveraging this geometrical dependence of $S/B$, SBND-PRISM will be less affected by the large cross section uncertainties on the background.

Now, we outline our statistical approach.
As we lack an understanding of systematic uncertainties that can lead to spectral distortions on neutrino energy distributions, and their correlations, we will do simple cut and count analyses.
We focus in two strategies: utilizing the total event rate at SBND and applying the SBND-PRISM concept.
The SBND sensitivity is evaluated through a $\chi^2$ test,
\begin{equation}
    \chi_{\rm SBND}^2 = \frac{N_{\nu_e, \rm sig}^2}{N_{\nu_e} + \sigma_{\nu_e}^2 N_{\nu_e}^2 }
    \label{eq:chi2_no_prism}
\end{equation}
where $N_{\nu_e, \rm sig}$ is the number of signal events, while $N_{\nu_e}$ is the number of intrinsic $\nu_e$ background events.
We take the systematic uncertainty on the background events to be  $\sigma_{\nu_e}=10\%$~\cite{MiniBooNE:2008hfu}.
The number of signal events coming from  nonstandard  decays of a meson $M=\pi,\,K$ has been calculated as
\begin{equation}
    N_{\nu_e, \rm sig} = N_{\nu_\mu}^{M}\, {\rm BR}(M^+\to\mu^+\nu_e),
\end{equation}
where $N_{\nu_\mu}^{M}$ is the standard prediction for the number of $\nu_\mu$ charged current events coming from parent meson $M$, as described in the previous section.
Note that this implicitly neglects differences in the CC cross sections for $\nu_e$ and $\nu_\mu$.
We will also estimate SBND's sensitivity to LNV decays, namely $\pi^+\to\mu^+\overline\nu_e$ and $K^+\to\mu^+\overline\nu_e$.
To that end we substitute the branching ratio appropriately in the definition of the signal, and we rescale the number of signal events by the ratio of $\overline\nu_e$ to $\nu_e$ cross sections.
Backgrounds are the same since LArTPCs cannot distinguish electrons from positrons.

For SBND-PRISM, we build a covariance matrix to correlate event rate in different layers as 
\begin{equation}
    C_{ij} =\left[(\sigma_{\nu_e}^{\rm c})^{2} + \delta_{i,j}(\sigma^{\rm u}_{\nu_e})^2 \right]N^i_{\nu_e}N^j_{\nu_e} + \delta_{i,j} N^{i}_{\nu_e},
\end{equation}
where $i,j$ denote the SBND-PRISM layer, $\sigma_{\nu_e}^{\rm c}$ and $\sigma^{\rm u}_{\nu_e}$ are the correlated and uncorrelated systematic uncertainties across layers, and $N_{\nu_e}^i$ is the number of intrinsic $\nu_e$ background events in layer $i$.
We take the correlated uncertainty to be $\sigma_{\nu_e}^{\rm c}=10\%$, consistently with the previous analysis, while we study two possibilities for the uncorrelated systematics, $\sigma^{\rm u}_{\nu_e}=5\%$ or $2\%$, corresponding to a pessimistic and an optimistic case.
We then build a $\chi^2$ as
\begin{equation}
    \chi^2_{\rm PRISM} = \sum_{i,j=1}^8 N_{\nu_e, \rm sig}^i (C^{-1})_{ij}N_{\nu_e, \rm sig}^j,
\end{equation}
where $i,j$ denote the SBND-PRISM layer and the number of signal events per layer is defined akin to what was described previously.
For both analyses we will calculate the  95\%~C.L. on the nonstandard branching ratio using $\chi^2=3.841$.

\section{Results and discussion}
\label{sec:results}

\begin{figure}[t]
            \centering
            \includegraphics[width = \textwidth]{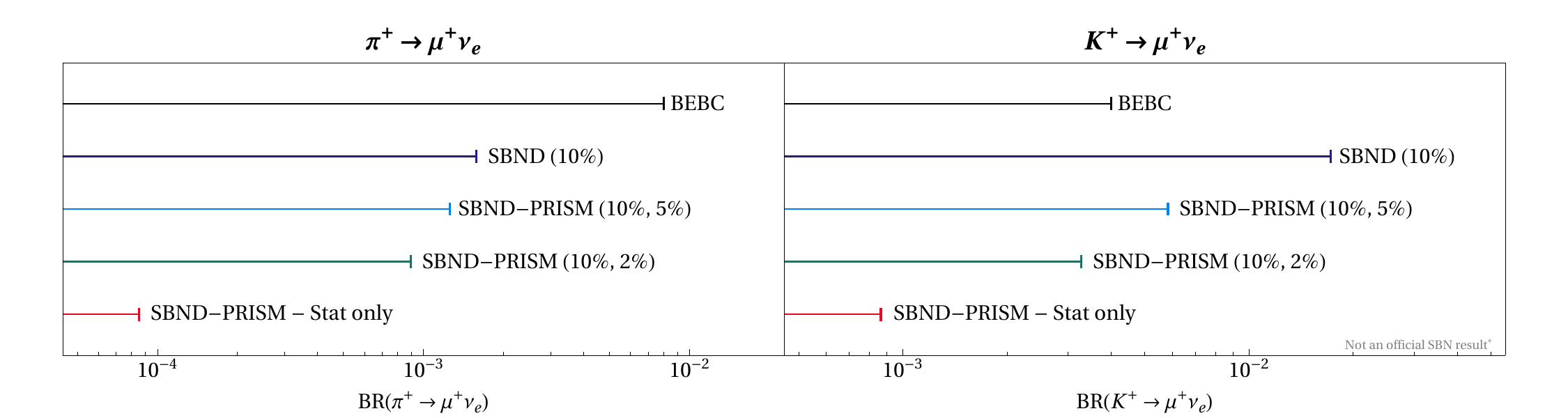}
            \caption{Sensitivity of SBND and SBND-PRISM to $\pi^+ \to \mu^+ \nu_e$ and $K^+ \to \mu^+ \nu_e$ lepton flavor violating decays. 
            For comparison, the BEBC~\cite{Cooper-Sarkar:1981bam, Lyons:1981xs, Workman:2022ynf} results and the statistics only SBND-PRISM limits are shown.}
            \label{fig:comparison_limits}
\end{figure}

We proceed to present our results for the SBND and SBND-PRISM sensitivity on nonstandard pion and kaon decays.
Starting with the lepton flavor violating decays, $\pi^+\to\mu^+\nu_e$ and $K^+\to\mu^+\nu_e$, we present in the left panel of \cref{fig:comparison_limits} and in \cref{tab:summary_results} the current BEBC constraints~\cite{Cooper-Sarkar:1981bam, Lyons:1981xs, Workman:2022ynf}, as well as the projected SBND sensitivities for an exposure of $10^{21}$ POT~\cite{palamara2023int} for three scenarios: SBND only with 10\% uncertainty; SBND-PRISM with 10\% (correlated) and 5\% (uncorrelated) uncertainties; and SBND-PRISM with 10\% (correlated) and 2\% (uncorrelated) uncertainties.
For reference, we also show the statistics only limit.
To facilitate the interpretation of this result in terms of new physics, we also show in \cref{tab:summary_results} a range of constraints on the scale of dimensions-6 operators that can lead to such nonstandard transitions.

We can see that, even without PRISM, SBND can improve the BEBC constraint on $\pi^+\to\mu^+\nu_e$ by a factor $\sim5$; while when employing the SBND-PRISM technique an improvement of $5-9$ can be achieved.
We can understand these improvements by noting a few differences between BEBC and SBND.
First, the number of background CC $\nu_e$ events in BEBC is a factor two larger than in SBND; namely,
$(\nu_e:\nu_\mu)^{\rm BEBC}=(30/1800)\sim 0.016$~\cite{Cooper-Sarkar:1981bam}, compared to  $(\nu_e:\nu_\mu)^{\rm SBND}= (3.3\times 10^{4}:4 \times 10^{6}) \sim 0.008$~\cite{MiniBooNE:2008hfu} for SBND. 
Besides that, BEBC had a much smaller statistical sample of events: $6060\pm440$ candidates $\nu_\mu$ CC events versus $\mathcal{O}(10^7)$ in SBND.
SBND-PRISM presents further improvement due to the change in the fractions of $\nu_e:\nu_\mu$ across different layers as discussed in \cref{sec:analysis}, see \cref{fig:evts_vs_bkg}.

For kaons, the situation is slightly different, as can be seen from the right panel of \cref{fig:comparison_limits} and \cref{tab:summary_results}.
Since BEBC was on the CERN SPS beamline, which accelerates protons to 450 GeV, the kaon production was significantly higher than in SBND which is in the 8 GeV Booster Neutrino Beam.
With a lower signal rate, the systematic uncertainties in the $\nu_e$ background become dominant in SBND, which limit the sensitivity to nonstandard kaon decays.
Nevertheless, SBND-PRISM can lead to improved constraints over BEBC, for the aggressive scenario with 2\% bin-to-bin uncorrelated systematics, by leveraging the larger asymmetry of muon neutrinos from kaons across each layer in comparison to the $\nu_e$ background.
This reduces the impact of systematic uncertainties, leading to a constraint that is a factor of $1.25$ better than BEBC's. 

For the lepton number violating decays, $\pi^+\to\mu^+\overline\nu_e$ and $K^+\to\mu^+\overline\nu_e$, we summarize our results in \cref{tab:summary_results}. 
For both pion and kaon LNV decays, our findings show that SBND does not improve the BEBC constraints, even in the optimistic SBND-PRISM scenario.
Although SBND will have higher statistics and smaller $\nu_e$ contamination than BEBC, the latter has an important advantage for this type of search.
BEBC, being a bubble chamber, can disentangle the $\nu_e$ and $\overline\nu_e$ backgrounds. The $\overline\nu_e$ contamination there is found to be below the $0.1\%$, leading to an essentially background free analysis.

\vspace{0.5cm}

Let us now discuss possible improvements to our proposed analysis.
First, we have not considered any spectral information due to the fact that we do not have access to realistic systematic uncertainties affecting the spectrum, particularly in the SBND-PRISM case where the correlations among spectra in different layers are relevant. 
Nevertheless, neutrinos coming from two-body $\pi$ and $K$ decays have very distinctive spectral features, the former being lower energy than the latter.
This features can be used to further improve background rejection, and most likely would lead to further improvement of all constraints presented here.
A handle that could be used to enhance the sensitivity to LNV decays is the event topology.
It has been observed that neutrino CC events tend to have more visible protons than antineutrino ones~\cite{Palamara:2016uqu}.
Zero-proton events are expected to have a better signal-to-background ratio than one-proton events, due to the smaller $\overline\nu_e$ contamination, which could ameliorate the LNV constraints in \cref{tab:summary_results}.
Last, a more direct way of distinguishing $\nu_e$ from $\overline\nu_e$ would be to magnetize liquid argon time projection chambers~\cite{snowmass_magnetize}.
This could also make feasible doing the LNV analysis when running in antineutrino mode.

\begin{table}[t]
    \renewcommand{\arraystretch}{1.2}
    \setlength{\tabcolsep}{10pt}
    \centering
    \begin{tabular}{|c|c|c|c|}
    \hline
    \multicolumn{4}{|c|}{Lepton Flavor Violating Decays} \\ \hline
    \hline
        Experiment &  $\rm{BR}(\pi^+ \to \mu^+ \nu_e)$ & $\rm{BR}(K^+ \to \mu^+ \nu_e)$ & $\Lambda^{(6)}$(TeV)  \\
        \hline
        
        BEBC & $8 \times 10^{-3}$ &  $4 \times 10^{-3}$ & 0.59 - 4.9\\           
           \hline
           SBND ($10\%$) & $1.5 \times 10^{-3}$ & $1.7 \times 10^{-2}$ & 0.89 - 4.6\\        
                \hline
                 SBND-PRISM ($10\%$, $5\%$) & $1.2 \times 10^{-3}$ & $5.8 \times 10^{-3}$ & 0.94 - 5.8\\
        \hline
                SBND-PRISM ($10\%$, $2\%$) & $8.9 \times 10^{-4}$ & $3.2 \times 10^{-3}$ & 1 - 6.8\\
        \hline
         Statistics only & $8.5 \times 10^{-5}$ & $8.6\times 10^{-4}$ & 1.8 - 9.4\\           
                \hline 
    \hline
    \multicolumn{4}{|c|}{Lepton Number Violating Decays} \\ \hline\hline
        Experiment & $\rm{BR}(\pi^+ \to \mu^+ \overline{\nu}_e)$ & $\rm{BR}(K^+ \to \mu^+  \overline{\nu}_e)$  & $\Lambda^{(6)}$ (TeV)\\
        \hline
         BEBC & $1.5 \times 10^{-3}$ & $3.3 \times 10^{-3}$ & 0.89 - 6.7\\           
           \hline
        SBND ($10\%$) &  $4.0 \times 10^{-3}$ & $3.9 \times 10^{-2}$ & 0.7 - 3.6\\        
                \hline
                                 SBND-PRISM ($10\%$,$5\%$)&  $3.1 \times 10^{-3}$& $1.3 \times 10^{-2}$ & 0.74 - 4.8 \\
        \hline
                 SBND-PRISM ($10\%$,$2\%$)&  $2.1 \times 10^{-3}$& $7.4 \times 10^{-3}$ & 0.82 - 5.5\\
        \hline
        Statistics only & $2.1 \times 10^{-4}$& $1.9 \times 10^{-3}$ & 1.5 - 7.7\\           
                \hline
    \end{tabular}
    \caption{Summary of the results for the pion and kaon LFV (upper part) and LNV (lower part) searches considering SBND and SBND-PRISM. 
    The values in parenthesis indicate the uncertainties (total flux for SBND, and correlated and uncorrelated for SBND-PRISM, respectively). 
    For comparison we also include the previous bounds from BEBC~\cite{Cooper-Sarkar:1981bam, Lyons:1981xs, Workman:2022ynf} and the SBND reach considering only statistical errors.
    We also include the range constraints that can be set on the scale of dimension-6 operators that can lead to those transitions.}
    \label{tab:summary_results}
\end{table}

\section{Conclusions}
\label{sec:conclusions}
In this work we have estimated how the Short-Baseline Near Detector can improve constraints on lepton flavor violating decays of pions and kaons.
In doing so, we have also shown how to leverage the Booster Neutrino Beam angular spread to mitigate the impact of systematic uncertainties --a technique known as SBND-PRISM~\cite{delTutto2021SBND}.
We find that SBND-PRISM can constrain 
${\rm BR}(\pi^+\to\mu^+\nu_e)<(0.89-1.2)\times 10^{-3}$ and 
${\rm BR}(K^+\to\mu^+\nu_e)<(3.2-5.8)\times 10^{-3}$, for aggressive and conservative assumptions on systematic uncertainties.
This represents improvement factors on these nonstandard decays between 6 to 9 for pion nonstandard decays and up to 1.25 for kaons with respect to previous bounds from the Big European Bubble Chamber.
We have also estimated SBND's sensitivity to lepton number violating decays 
$\pi^+(K^+)\to\mu^+\overline\nu_e$.
However, we have not found any improvement over previous BEBC constraints due larger backgrounds in SBND.
We have identified several ways this analysis may be further improved by the SBND collaboration, including leveraging spectral information and event topology.
The interesting possibility of magnetizing LArTPCs would also bolster the experiment sensitivity to LNV decays.

\section*{Acknowledgement}
We thank Gustavo Burdman, Marco del Tutto, Ornella Palamara and Jure Zupan for invaluable discussions.
G.F.S.A. is fully financially supported by Fundação de Amparo à Pesquisa do Estado de São Paulo (FAPESP) under Contracts No. 2022/10894-8 and No. 2020/08096-0. G.F.S.A. would like to thank the hospitality of the Fermilab Theory Group. R. Z. F. is partially supported by FAPESP under Contract No. 2019/04837-9, and by  Conselho Nacional de Desenvolvimento Científico e Tecnológico (CNPq).
P.M. is supported by Fermi Research Alliance, LLC under Contract No. DE-AC02-07CH11359 with the U.S.
Department of Energy, Office of Science, Office of High Energy Physics.

\bibliography{pion}
\end{document}